\documentclass[a4paper,fleqn,usenatbib]{mnras}
\usepackage[T1]{fontenc}
\usepackage{aecompl}

\usepackage{newtxtext,newtxmath}

\usepackage[T1]{fontenc}
\usepackage{ae,aecompl}

\usepackage{graphicx}   
\usepackage{amsmath}    
\usepackage{amssymb}    

\title[The NuSTAR observation of Ark~564]{The high-Eddington NLS1 Ark~564 has the coolest corona}
\author[Kara et al.]{E. Kara$^{1,2}$\thanks{E-mail:
ekara@astro.umd.edu}, J. A. Garc\'ia$^{3,4,5}$, A. Lohfink$^{6}$, A. C. Fabian$^{6}$, C. S. Reynolds$^{1}$,
\newauthor
F. Tombesi$^{2,1}$ and D. R. Wilkins$^{7}$\\
$^{1}$Department of Astronomy, University of Maryland, College Park, MD 20742\\
$^{2}$X-ray Astrophysics Laboratory, NASA/Goddard Space Flight Center, Greenbelt, MD 20771\\
$^{3}$Cahill Center for Astronomy and Astrophysics, California Institute of Technology, Pasadena, CA 91125\\
$^{4}$Remeis Observatory \& ECAP, Universität Erlangen-Nürnberg, Sternwartstrasse 7, D-96049 Bamberg, Germany\\
$^{5}$Harvard-Smithsonian Center for Astrophysics, 60 Garden Street, Cambridge, MA 02138, USA\\ 
$^{6}$Institute of Astronomy, Madingley Rd, Cambridge CB3 0HA, United Kingdom\\
$^{7}$KIPAC, Stanford University, 452 Lomita Mall, Stanford, CA 94305\\
}
\begin{document}

\date{\today}

\pagerange{\pageref{firstpage}--\pageref{lastpage}} \pubyear{2015}

\maketitle

\label{firstpage}

\begin{abstract}
Ark~564 is an archetypal Narrow line Seyfert 1 that has been well observed in soft X-rays from 0.3--10~keV, revealing a steep spectrum, strong soft excess, iron~K emission line and dramatic variability on the order of hours.  Because of its very steep spectrum, observations of the source above 10~keV have been sparse. We report here on the first {\em NuSTAR} observation of Ark~564. The source was observed for 200~ks with {\em NuSTAR}, 50~ks of which were concurrent with {\em Suzaku} observations. {\em NuSTAR} and {\em Suzaku} observed a dramatic flare, in which the hard emission is clearly delayed with respect to the soft emission, consistent with previous detections of a low-frequency hard lag found in {\em XMM-Newton} data.  The {\em NuSTAR} spectrum is well described by a low-temperature Comptonisation continuum (with an electron temperature of $15\pm2$~keV), which irradiates a highly ionised disc. No further relativistic broadening or ionized absorption is required.  These spectral results show that Ark~564 has one of the lowest temperature coronae observed by {\em NuSTAR} to date.  We discuss possible reasons for low-temperature coronae in high-Eddington sources.
\end{abstract}


\section{Introduction}

Narrow-line Seyfert 1 galaxies (NLS1s) were initially classified for their striking optical properties--narrow H$\beta$ lines, weak [OIII] and strong FeII lines \citep{osterbrock85}, which suggested these Active Galactic Nuclei (AGN) contained relatively small central supermassive black holes accreting near or even above the Eddington limit \citep{boroson92}.  In addition to their unique optical spectra, NLS1s show interesting behaviour in the X-ray band, including steep 2--10~keV photon indices \citep{brandt97} and very rapid variability, which corroborates the picture of low-mass black holes accreting at high-Eddington rates, where the X-ray emitting corona is cooled via inverse Compton scattering by the intense UV flux of the accretion disc (e.g. \citealt{pounds95}).  Such a model predicts that, for a given optical depth, the temperature of the corona in steep spectra NLS1s will be cooler than in other Seyferts (\citealt{haardt93,zdziarski94}). However due to the intrinsically steep spectra of these sources and the low-sensitivity of previous hard X-ray instruments, observations of steep spectrum NLS1s above 10~keV have been sparse. In this paper, we present the first results of a deep observation of the well-known NLS1 Ark~564 taken with the {\em Nuclear Spectroscopic Telescope Array} ({\em NuSTAR}; \citealt{harrison13}). We find  one of the lowest temperature coronae discovered with {\em NuSTAR}.   

Apart from the coronal continuum, both broad and narrow-line Seyfert 1s show interesting spectral features, including a soft excess below 1~keV and a dip at $\sim 7$~keV. 
Since their discoveries, the soft excess and iron~K feature have been subjects of debate.  The soft excess is a largely featureless component that can be well-modelled by an additional Comptonisation component with a temperature of 0.2~keV. However, this temperature is constant, regardless of the mass or mass accretion rate of the central black hole, which requires some fine-tuning (see \citealt{czerny03,gierlinski04}). Alternatively, the soft excess can be well described by a reflection model where the corona irradiates a partially ionised disc, causing the fluorescence of atomic lines that are relativistically smeared due to the proximity to the central black hole \cite{fabian89}. In this reflection model, the dip at 7~keV is interpreted as the blue wing of relativistically broadened iron~K$\alpha$ emission line.  Alternatively, the strong dip (especially in the extreme NLS1 1H0707-495) has been interpreted as a P-Cygni profile from emission and absorption in a Compton-thick wind \citep{done07}, or even completely via absorption through distant partial covering clouds \citep{mizumoto14}, though this model requires absorption from several different ionization zones with the same covering factor (see \citealt{kara15b} for more details).

The basic picture of reflection off the inner accretion disc is supported by the recent discovery of X-ray reverberation, where the soft excess and iron~K band are seen to respond seconds to minutes after the X-ray continuum, due to the additional light path these photons traveled. These time lags were initially found between the strong soft excess and the X-ray continuum in several variable NLS1s (e.g. 1H0707-495, \citealt{fabian09,zoghbi10}, and tentatively in Ark~564, \citealt{mchardy07}), and later in several other Seyfert~1s (e.g. \citealt{demarco13}).  Reverberation was later found associated with the broad iron~K emission line \citep{zoghbi12,zoghbi13,kara13c,kara16}. Through a recent study of a large sample of variable Seyfert galaxies, we find that reverberation is common in both broad and narrow-line Seyfert~1 galaxies, and find tentative evidence that the inferred distance from the corona to the disc increases with mass accretion rate \citep{kara16}.    

Further support of the ionised reflection model has come from observations above 10~keV, which can now be probed thanks to the unprecedented hard-band sensitivity of {\em NuSTAR}.  In its first four years in orbit, {\em NuSTAR} has observed several broad line Seyfert~1 galaxies (e.g. NGC~1365, \citealt{risaliti13,walton13}, Mrk~335; \citealt{parker14,wilkins15}, Fairall~9; \citealt{lohfink16}) and a few NLS1s (Swift~J2127.4+5654; \citealt{marinucci14a}, 1H0707-495; \citealt{kara15b}), and finds a ubiquitous hard excess above the continuum, peaking at $\sim 20-30$~keV.  This feature is predicted by reflection scenarios as coronal photons above 10~keV will Compton scatter off the accretion disc, and also rules out models that try to explain the spectrum through absorption alone.  Furthermore, the so-called Compton hump has also been seen to reverberate a few hundred seconds after the continuum emission in three Seyferts \citep{zoghbi14,kara15a}.

In addition to measuring accretion disc parameters from observations of reflection, {\em NuSTAR} can also put strong constraints on the temperature of the continuum-emitting corona by measuring the high-energy cut-off.  Cut-offs have been measured in a few AGN thus far \citep{balokovic14, brenneman14, marinucci14a,tortosa16}, and constraining upper limits have been found in several others (see \citealt{marinucci16} for a compilation).  Recently, \citet{fabian15} found that most of the coronae observed by {\em NuSTAR} thus far have temperatures that lay close to the physically allowed limit.  In a compact corona, where a significant fraction of the upscattered X-ray luminosity is radiated at energies above 511 keV (the electron rest energy), many of the high-energy photons will not escape the corona, but instead will collide with other high-energy photons, producing electron-positron pairs. Therefore, any additional heat to the corona will not contribute to more high-energy radiation above 511 keV, but rather will create more electron-positron pairs at lower energy, thus putting a hard limit on the temperature of the corona \citep{svensson82, guilbert83}. The results of \citet{fabian15} suggest that many of the coronae observed thus far could be pair-dominated plasmas.

We expand upon these spectral and timing studies with {\em NuSTAR} by observing one of the X-ray brightest NLS1 galaxies, Ark~564. While two other steep NLS1s have been observed with {\em NuSTAR} (1H0707-495 and IRAS~13224-3809), Ark~564 is almost an order of magnitude brighter, and so we can now place interesting constraints on the Compton hump and high-energy cut-off in a steep NLS1. 

\subsection{The X-ray bright NLS1 Ark~564}

Ark~564 ($z=0.02468$) is a popular source for X-ray studies because it exemplifies the unusual characteristics of a NLS1: steep X-ray spectrum, strong soft excess, and rapid variability. It also has the benefit of being extremely bright in the soft X-ray band ($F_{0.3-10~\mathrm{keV}}=1.4\times10^{-10}$~erg/s/cm$^2$  during this campaign). Because of these qualities, it has been observed by all the major X-ray observatories, and now we include {\em NuSTAR} to this list.

\begin{figure*}
\includegraphics[width=\textwidth]{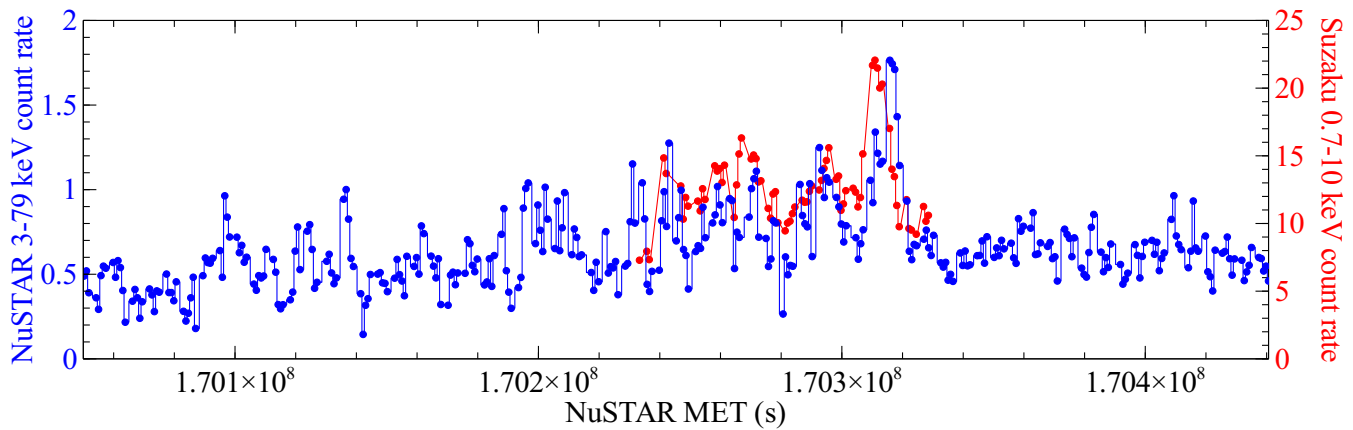}\\
\caption{{\em NuSTAR} FPMA+FPMB (blue; left axis) and {\em Suzaku} XIS FI+BI (red; right axis) light curves in 800~s bins. The {\em NuSTAR} exposure is 200~ks. Suzaku joined the campaign for a 50~ks exposure and fortuitously observed a bright flare.  There is a clear time delay between the {\em Suzaku} and {\em NuSTAR} flares. See Fig.~\ref{lc_evolution} for a more detailed look at the flare in different energy bands.} 
\label{lc}
\end{figure*}

\begin{figure*}
\includegraphics[width=\textwidth]{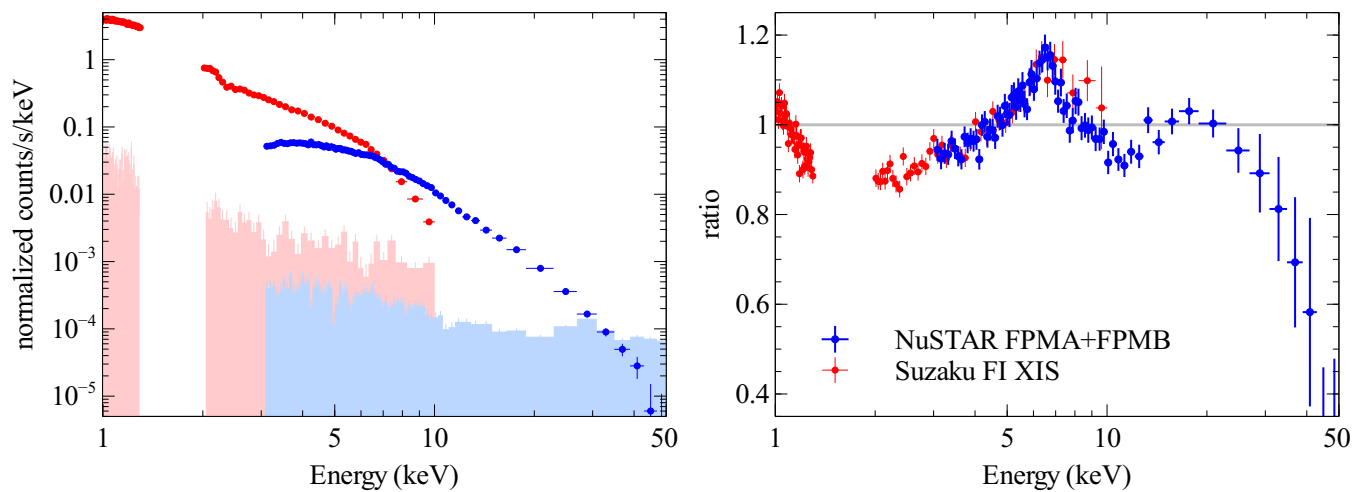}\\
\caption{{\em left}: {\em NuSTAR} FPMA+FPMB (blue) and {\em Suzaku} XIS FI (red) spectra and their respective background spectra (filled regions). {\em right:} The {\em NuSTAR} and {\em Suzaku} spectra fit to a simple powerlaw with Galactic absorption. The photon index and normalization was free to vary as the {\em Suzaku} observation occurred during a higher flux epoch. The spectrum is steep ($\Gamma \sim 2.5$) and there is a clear iron~K line and fall-off at high energies.} 
\label{spec}
\end{figure*}

In addition to its broadband spectral features, Ark~564 has been studied with high-resolution spectrometers, and a low-ionization warm absorber is found (e.g. \citealt{giustini15,laha14,tombesi10,papadakis07}).  Interestingly, \citet{giustini15} and most recently \citet{khanna16} found through long {\em XMM-Newton}-RGS observations that the low ionization warm absorber is unusually low-velocity compared to other Seyferts.  While these studies are important for obtaining a complete picture of the AGN, with the limited soft-band coverage we have in these current observations (50~ks with {\em Suzaku}-XIS), we do not require any additional warm absorber component to describe the data.

Ark~564 is also a very unusual source in terms of its timing properties.  Eight years of {\em RXTE} observations, together with {\em ASCA} and {\em XMM-Newton} observations allowed for measuring the X-ray power spectral density (PSD) over 8 decades in temporal frequency \citep{mchardy07}.  This revealed the presence of a low-frequency break in the PSD, which had never been seen in other Seyfert galaxies (e.g. \citealt{uttley02}, \citealt{markowitz03}). The authors noted that the PSD of Ark~564 is very reminiscent of the X-ray binary Cyg X-1 in the luminous hard state, and suggested that Ark~564 is the AGN analog of this accretion state (see also \citealt{arevalo06b} and Section~\ref{discuss} in this paper).  


The rapid variability of Ark~564 has also revealed interesting physics.  In \citet{kara13c}, we used a 500~ks {\em XMM-Newton} of Ark~564 \citep{legg12} to measured the lag-energy spectrum at different temporal frequencies.  We found that the low-frequency lags ($\sim 10^{-5}-10^{-4}$~Hz) show a featureless log-linear increase with energy (consistent with the lag being associated with the continuum; \citealt{kotov01}, \citealt{arevalo06}), and that the high-frequency lags ($\sim 10^{-4}-10^{-3}$~Hz) show clear evidence for iron~K reverberation. This confirmed that two separate processes were responsible for the low- and high-frequency lags, and that the high-frequency lags were not a phase wrapping artefact from reverberation of distant circumnuclear material (e.g. \citealt{miller10b}). 

In this paper we extend the X-ray observations of Ark~564 to high energies through a 200~ks observation with the {\em NuSTAR} observatory. The paper is organized as follows: in Section~\ref{obs} we describe the observations and data reduction. In Section~\ref{results} we present the results of the timing and spectral fitting analyses, and discuss the interpretations of these results in Section~\ref{discuss}. Unless otherwise specified for one interesting parameter, the error bars indicate the 90\% confidence interval. Throughout this paper, luminosities are calculated using a $\Lambda$CDM cosmological model with $H_0=71$~km~s$^{-1}$~Mpc$^{-1}$.

\section{Observations}
\label{obs}

{\em NuSTAR} observations of Ark~564 were taken on 2015-May-22 for a total exposure of 200~ks. During the observation, the AGN was variable (as usual), but there was also a particularly striking flare, where the 3--79~keV flux increased by a factor of 3 for $\sim 5$~ks (Fig.~\ref{lc}; blue points). The {\em NuSTAR} observations were accompanied by {\em Suzaku} observations starting on 2015-May-25 for 50~ks in order to obtain some simultaneous soft X-ray coverage. Fortuitously, the 50~ks {\em Suzaku} exposure occurred during the time of the flare (Fig.~\ref{lc}; red points). There is a clear delay between the flares as seen by {\em Suzaku} and {\em NuSTAR}.  We describe this flare in more detail in Section~\ref{timing_results} on the timing analysis results.
 
The {\em NuSTAR} Level 1 data products were processed with the NuSTAR Data Analysis Software (NUSTARDAS v1.6.0), and the cleaned Level 2 event files were produced and calibrated with the standard filtering criteria using the NUPIPELINE task and CALDB version 20160502. We extracted several region sizes in order to obtain the highest signal-to-noise spectra, and found the optimal source region was a circular source region of radius 50 arcsec. We used a background region of radius 60 arcsec. The regions were the same for both instruments, Focal Plane Module A and Focal Plane Module B (FPMA/FPMB). The spectra were binned in order to oversample the instrumental resolution by a factor of 2.5 and to have a signal-to-noise ratio of greater than $3\sigma$ in each spectral bin, though coarser binning oversampling the spectral resolution to a factor of at least 3 and to a signal-to-noise of $>5\sigma$ resulted in very similar best-fit parameters.

\begin{figure}
\includegraphics[width=\columnwidth]{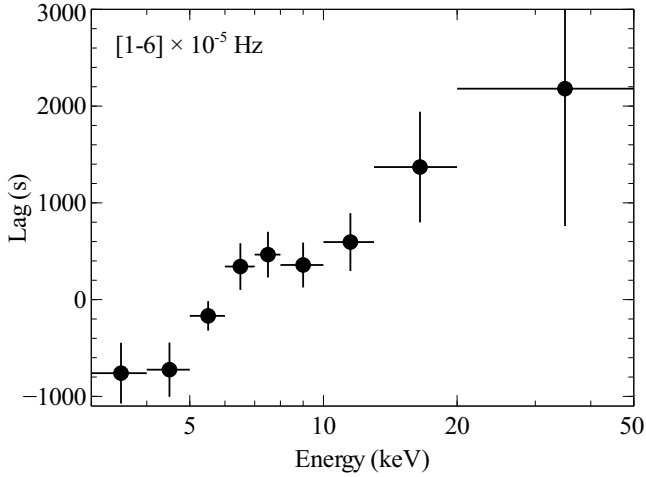}\\
\caption{Frequency-resolved time lags as a function of energy at frequencies [$1-6]\times 10^{-5}$~Hz. The lag is calculated between each energy bin and the entire reference band from 3--50~keV (excluding the bin-of-interest to avoid correlated errors). The error bars indicate $1\sigma$ confidence intervals. These frequency-resolved time lags show qualitatively similar results to Fig.~\ref{lc_evolution}, where the time lag is seen directly in the light curve.}
\label{lagen}
\end{figure}

The {\em Suzaku} observations were processed from the unfiltered event files for each of the XIS CCDs.  The source and background regions were $3.5$ arcmin in radius. {\sc XSELECT} was used to extract the spectral products, and the responses were generated with medium resolution using {\sc XISRESP}.  The front-illuminated (FI) spectra and responses were combined.  The back-illuminated (BI) spectrum was noisy, and so it was not included in the spectral modelling.  However, it was checked for consistency with the front-illuminated spectrum.  The XIS spectrum was binned to a minimum of 25 counts per bin. We visually inspected the band from 1-2.5~keV, where there are known calibration uncertainties, and found two sharp instrumental features around 1.4 and 1.8~keV, and therefore we excised the 1.3-2~keV band.  The spectral fitting was limited to data above 1~keV in order to avoid the low energy bands that were known to be effected by build-up of molecular contamination on the optical block filter, especially at the end of the {\em Suzaku} mission\footnote{http://space.mit.edu/XIS/monitor/contam/}. For the timing analysis, we used the 0.7--10~keV band, since the calibration effects do not effect the relative flux at each time bin.

\begin{figure}
\begin{center}
\includegraphics[width=\columnwidth]{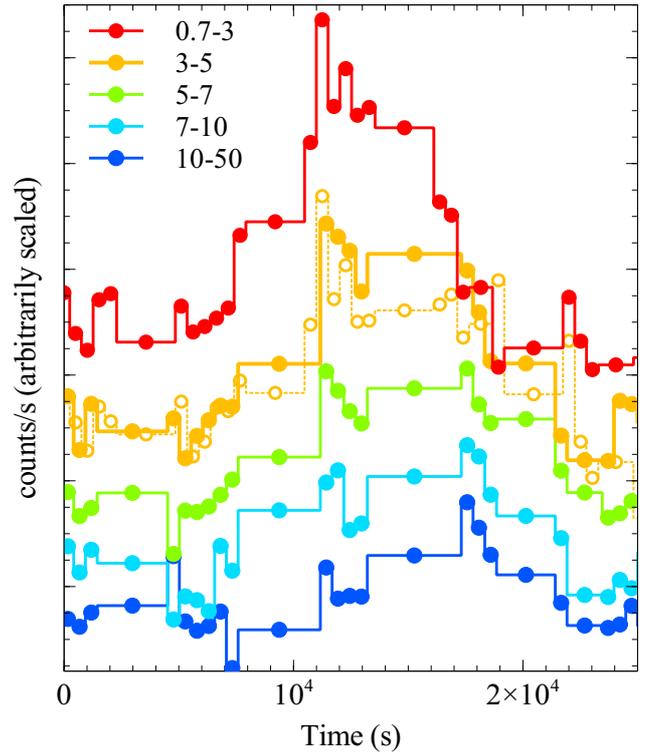}\\
\caption{A more detailed look at the flare shown in Fig.~\ref{lc}. The start time on the x-axis corresponds to $1.703\times10^8$~seconds MET, and the y-axis is the light curve count rate scaled arbitrarily for visual purposes. We use 512~s time bins for both {\em Suzaku} and {\em NuSTAR} light curves, with the larger bins being due to the orbital gaps of both instruments.  The light curves increase in energy from top to bottom, and the exact energy ranges in units of keV are shown in the figure key. We show both {\em Suzaku} and {\em NuSTAR} light curves of the 3--5~keV band for comparison (open circles for {\em Suzaku}, filled for {\em NuSTAR}).  The peak of the flare in the 10--50~keV band is roughly 6000 seconds after the peak of the flare in the 0.7--3~keV band. This is consistent with the amplitude of the hard lag found in the Fourier analysis, shown in Fig~\ref{lagen}.} 
\label{lc_evolution}
\end{center}
\end{figure}

Fig.~\ref{spec} shows the {\em NuSTAR} (FPMA/FPMB combined) and {\em Suzaku}-XIS (FI) spectra and the respective background spectra for each instrument. The {\em NuSTAR} spectrum dominates over the noise up to 27~keV.  The panel on the right shows the {\em Suzaku} and {\em NuSTAR} spectra fit to a simple powerlaw with Galactic absorption ($n_{\mathrm{H,Gal}}=5.3\times 10^{20}$~cm$^{-2}$; Leiden/Argentine/Bonn Survey of Galactic HI; \citealt{kalberla05}). The photon index and normalization were left to vary to account for possible spectral variability during the flaring epoch. Ark~564 has a strong soft excess and a clear iron line peaking at 6--7~keV. The high-energy coverage of {\em NuSTAR} reveals for the first time that the edge of the iron line extends to $\sim 11$~keV. Fitting the 3--10~keV  {\em NuSTAR} spectrum with a powerlaw plus Gaussian reveals an equivalent width of the iron line of $280\pm50$ eV. This is much stronger than previous detections of the iron~K line with {\em XMM-Newton} (e.g. \citealt{papadakis07},\citealt{giustini15}), which found equivalent widths of $\sim 100$~eV. The smaller equivalent width found in {\em XMM-Newton} observations is possibly an effect of residual pile-up, as the only inconsistency between the {\em XMM-Newton} and {\em NuSTAR} spectra is a $\sim 15-20\%$ excess in the 7--10~keV {\em XMM-Newton} spectrum with respect to that from {\em NuSTAR}. 

These {\em NuSTAR} observations reveal the first clear detection of the iron line band and a hard excess above 10~keV (often interpreted as the Compton reflection hump). Also evident in the spectrum are hints of a  downturn starting at 20~keV. In Section~\ref{spec_results}, we report on results of the spectral modelling of these {\em Suzaku} and {\em NuSTAR} spectra.

\section{Results}
\label{results}

\subsection{Timing results}
\label{timing_results}

Ark~564 is highly variable in the soft X-rays and therefore it has been the target for several time lag studies. 
Now with these {\em NuSTAR} observations, we can attempt to extend the timing analysis up to energies above 10~keV. Time lag analysis of other AGN observed with {\em NuSTAR} has been very elucidating for high-frequency reverberation studies, revealing both the iron~K and Compton hump lag in MCG-5-23-16 \citep{zoghbi14} and Swift~J2127.4+5654 \citep{kara15a}.  To perform a frequency-resolved time lag analysis on frequencies longer than the {\em NuSTAR} orbital frequency ($\sim 10^{-3}$~Hz), we use the Maximum likelihood technique described in \citet{zoghbi13b}.  This is in essence a time-domain approach, which directly fits the light curves for the most likely frequency-dependent phase lag \citep{miller10b}.  The errors represent the 68\% confidence intervals on the values that change -2log$(\mathcal{L}/\mathcal{L}_{\mathrm{max}})$ by 1 (see \citealt{zoghbi13b} for more details).  

In Fig.~\ref{lagen}, we show the lag-energy spectrum at frequencies $[1-6] \times 10^{-5}$~Hz computed using the Maximum likelihood method.  The lag-energy spectrum shows the time lag between each bin of interest and a broad reference band, which we take to be the 3--50~keV band in this case (with the bin-of-interest removed).  As the reference band is nearly the same for each bin-of-interest, the meaningful time lag is the relative time lag between each small bin.  We see the lag increases steadily with energy, similar to the results found in \citet{arevalo06b} and \citet{kara13c} using {\em XMM-Newton} data. Unfortunately, due to limited statistics we were unable to probe higher frequencies, where iron~K reverberation was found previously.

\begin{figure}
\includegraphics[width=\columnwidth]{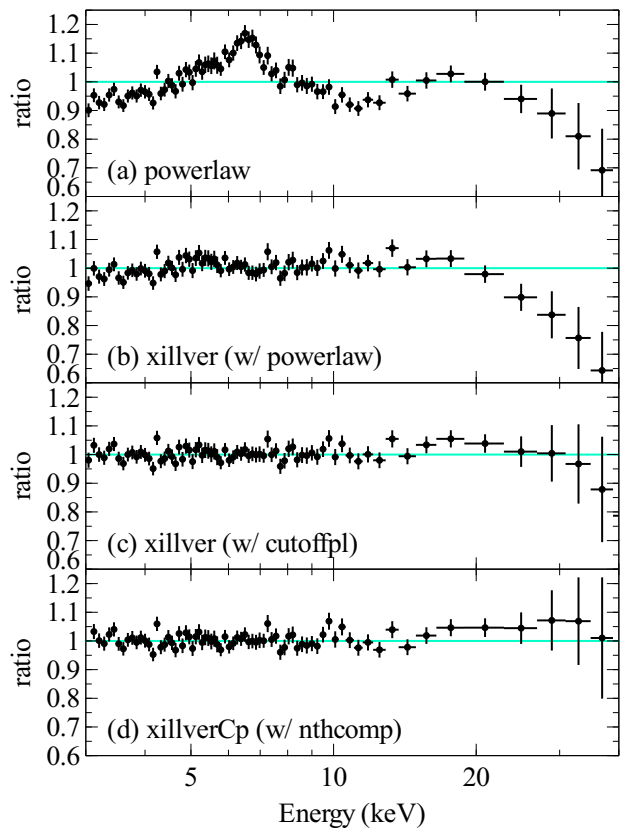}\\
\caption{The ratio of the full FPMA+FPMB {\em NuSTAR} spectra to different models. (a) The ratio plot to a simple powerlaw. (b) The ratio plot to an ionised reflection model with no high-energy cut-off. (c) The ratio plot to an ionised reflection model with high-energy cut-off.  The best fit cut-off energy using the {\sc xillver} model is $E_{\mathrm{cut}}=42\pm3$~keV. (d) The ratio plot to {\sc xillverCp}, an ionised reflection model irradiated by a Comptonization continuum ({\sc nthcomp}; rather than a cutoff powerlaw). The best fit coronal electron temperature using the {\sc xillverCp} model is $T_{\mathrm{e}}=15\pm2$~keV. The details of the parameters for the best fit {\sc xillver} and {\sc xillverCp} can be found in the first two columns of Table~\ref{spec_params}.}
\label{ratio}
\end{figure}

\begin{figure}
\includegraphics[width=\columnwidth]{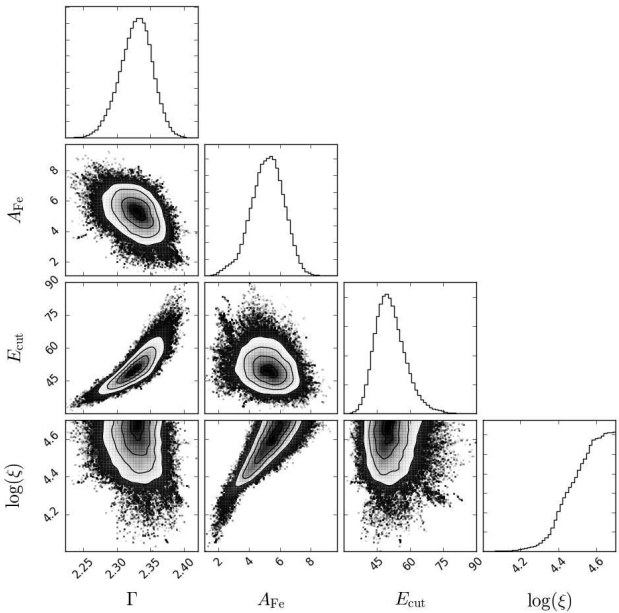}\\
\caption{Results of the MCMC analysis of the ionised reflection model applied to the {\em NuSTAR} data. We show the outputs for photon index $\Gamma$, iron abundance $A_{\mathrm{Fe}}$, cut-off energy $E_{\mathrm{cut}}$ (keV) and ionization parameter log($\xi$) (log(erg cm s$^{-1}$)). The 1D histograms show the probability distribution for each parameter normalized to unity.}
\label{corner}
\end{figure}

In addition to finding hard lags using a sophisticated frequency-resolved Maximum Likelihood approach, we can also see the hard lag by eye during the bright flare. Fig.~\ref{lc_evolution} shows the {\em Suzaku} and {\em NuSTAR} light curves in several energy bands from softest (top; red) to hardest (bottom; blue).  We zero-in on the time just around the flare ($t_0=1.703 \times 10^{8}$~s for comparison to Fig.~\ref{lc}). The peak of the flare is around $t_0+10^{4}$~s in the 0.7--3~keV band and by the 10--50~keV band, the peak has shifted by several thousands of seconds.  The amplitude of the lag found during this short flare is consistent with what we find from the frequency-resolved timing analysis performed on the entire light curve (Fig.~\ref{lagen}), and is similar in energy-dependence and timescale to the low-frequency lags found in archival {\em XMM-Newton} data \citep{legg12,kara13c}.

\subsection{Spectral results}
\label{spec_results}

\subsubsection{{\em NuSTAR} fits}

\begin{table*}
\centering
\begin{tabular}{l|ll|ll}
\hline
& \multicolumn{2}{c}{{\em NuSTAR}} & \multicolumn{2}{c}{{\em NuSTAR}+{\em Suzaku}}\\
\hline
& {\sc Xillver} & {\sc XillverCp} & {\sc Bremss + Relxilllp} & {\sc Xstar(Bremms + Xillver)}\\
$n_{\mathrm{H}} \times 10^{24}$ (cm$^{-2}$) & -- & -- & -- & $1.4^{+0.7}_{-0.6}$ \\
log($\xi_{\mathrm{wind}}$) (log(erg cm s$^{-1}$) & -- & -- & -- & $3.7\pm 0.1$ \\ 
$z_{\mathrm{wind}}$ & --  & -- & -- & $0.413\pm0.02$ \\
$h$ ($GM/c^2$) & -- & -- &$2.6^{+47}_{-0.03}$ & -- \\
$a$ &-- & -- & $0^{+0.99}_{-0.}$ & -- \\
$i$ (deg) & $22^{+7}_{-2}$ & $45\pm45$ & $63\pm6$ & $45\pm45$ \\
$\Gamma$ & $2.27\pm0.08$ & $2.32^{+0.02}_{-0.01}$ & $2.54^{+0.04}_{-0.09}$ & $2.389\pm0.004$\\
log($\xi$) (log(erg cm s$^{-1}$) & $>4.3$ & $4.43\pm0.06$ & $4.02^{+0.04}_{-0.06}$ & $4.6\pm0.1$ \\
$A_{\mathrm{Fe}}$ & $5.8^{+3.0}_{-1.9}$ & $1.4\pm0.2$ & $3.6^{+0.9}_{-2.85}$ & $1.1\pm0.3$ \\
$E_{\mathrm{cut}}$ (keV) & $42\pm3$ & -- & $63^{+4}_{-2}$ & $46\pm3$ \\
$T_{\mathrm{e}}$ (keV) & -- & $15^{+2}_{-1}$ & -- & -- \\
$R$ & $1.19^{+139}_{-0.8}$ & $1.18^{+2.1}_{-1.18}$ & $0.7^{+0.37}_{-0.5}$ & $1.0\pm0.6$ \\
$A_{\mathrm{refl}} \times 10^{-4}$ & $0.26^{+0.04}_{-0.15}$ & $0.24^{+0.3}_{-0.06}$ & $3\pm1$  & $1.4\pm1$ \\
$kT$ (keV) & -- & -- & $0.19^{+0.06}_{-0.03}$  & $0.25^{+0.06}_{0.04}$ \\
$A_{\mathrm{Bremss}}$ & -- & -- & $0.5^{+1.0}_{-0.3}$ & $0.43\pm0.2$ \\
\hline
$\chi^2/{\mathrm{d.o.f.}}$ & 378/374 & 380/374 &1248/1235 & 1246/1234\\
\hline
\end{tabular}
\caption{Parameters of the best fit spectral models for two models fit to the {\em NuSTAR} data alone and two models fit to the combined {\em NuSTAR}+{\em Suzaku} spectra. In order from top to bottom, the parameters are: the column density of the wind ($n_{\mathrm{H}}$), log of the ionization parameter of the wind ($\xi_{\mathrm{wind}}$), blueshift on the absorption feature ($z_{\mathrm{wind}}$), height of the continuum point source ($h$), inclination of the disc ($i$), photon index ($\Gamma$), log of the disc ionization parameter ($\xi$), iron abundance of reflector ($A_{\mathrm{Fe}}$), continuum cut-off energy ($E_{\mathrm{cut}}$), coronal temperature ($T_{\mathrm{e}}$), reflection fraction ($R$), normalization of the reflection component ($A_{\mathrm{refl}}$), and the plasma temperature ($kT$) and normalization ($A_{\mathrm{Bremss}}$) of thermal bremsstrahlung component.  The fits, in order from left to right show: {\em NuSTAR} alone fit with a powerlaw with exponential cut-off and ionized reflection ({\sc xillver}), {\em NuSTAR} alone fit with a physical Comptonization model and ionized reflection {\sc xillverCp}, {\em NuSTAR+Suzaku} fit with {\sc relxilllp} and an additional Bremsstrahlung component, and finally {\em NuSTAR+Suzaku} fit with ionised reflection and Bremsstrahlung component through an ionized outflowing wind.}
\label{spec_params}
\end{table*}

Ark~564 shows similar spectral properties to other well-known extreme Narrow-line Seyfert I galaxies, like 1H0707-495 and IRAS~13224-3809. It shows a strong soft excess, strong reprocessing features and a very steep spectrum.  The benefit of Ark~564 is that it is ten times brighter in soft X-rays than either of these sources, and therefore, it provides us with the best opportunity for studying the high-energy emission in a very steep spectrum NLS1.

We begin our fitting procedure by examining the 3--50~keV {\em NuSTAR} spectrum (Fig.~\ref{ratio}).  The ratio to a simple powerlaw model (Fig.~\ref{ratio}-a) shows a strong iron~K emission line, and so we start by modelling this with an ionised reflection model with cut-off energy fixed to the maximum value to see if the curvature of the high-energy emission can be described by the Compton reflection hump ({\sc xillver}; \citealt{garcia13}). We obtain a fairly good description below 10~keV, but find strong residual curvature above 10~keV (Fig.~\ref{ratio}-b).  Therefore, we allow the high-energy cut-off to vary, and find a cut-off is constrained to be $42\pm3$~keV (Fig.~\ref{ratio}-c, and best fit parameters in Column~1 of Table~\ref{spec_params}).  Allowing the cut-off to vary does not significantly change any of the other reflection parameters.

To examine the dependences on the fitting parameters, we performed Markov Chain Monte Carlo (MCMC) analysis, similar to \citet{parker15}, using the {\sc XSPEC\_EMCEE} code by Jeremy Sanders\footnote{available on github, https://github.com/jeremysanders/xspec\_emcee}. This is a script to use emcee (a pure-Python implementation of of Goodman \& Weare's Affine Invariant MCMC Ensemble sampler) to analyze X-ray spectra in {\sc xspec}. We use 50 walkers with 10,000 iterations each, burning the first 1,000. The walkers started at the best fit values found in {\sc xspec}, following a Gaussian distribution in each parameter, with the standard deviation set to the delta value of that parameter. The MCMC results of the ionised reflection model applied to the {\em NuSTAR} data are shown in Fig.~\ref{corner} for four relevant parameters: photon index, iron abundance, cut-off energy and ionization parameter. The ionization parameter is pegged to its maximum value, and there is a degeneracy between this parameter and the iron abundance (higher ionization parameter requires a higher iron abundance).  The photon index and cut-off energy are both tightly constrained, though there is a slight degeneracy between these parameters.

The ionised reflection model is driven to extremely high ionisation parameters in order to fit the feature at 11~keV, which is produced by the iron~K edge near 7-8~keV and the continuum photoelectric absorption of iron, both smeared by multiple scatterings in the hotter region of the disc's atmosphere. The resulting feature occurs at higher energies than other Seyfert galaxy observed with {\em NuSTAR} (see a sample of {\em NuSTAR} reflection spectra in Fig.~1 of \citealt{fabian14}).  The high ionisation smears the reflection spectrum, and we find that high ionisation alone is enough to broaden the iron line in these data (e.g. additional relativistic broadening is not statistically required).  As the reflecting surface becomes more ionised, the reflection spectrum begins to resemble the irradiating continuum. To counteract this, the iron abundance is driven to super-solar values to fit the very clear iron line in the spectrum (hence the degeneracy between ionization parameter and iron abundance). 

We still obtain a fairly good fit if we require the iron abundance to be fixed to solar value ($\chi^2/{\mathrm{dof}=405/375=1.08}$), though the super solar iron abundance is preferred at $>4\sigma$ confidence.  The result of fixing the iron abundance to solar values is that the ionisation parameter decreases, causing stronger residuals at $\sim 11$~keV. We also attempted fits including neutral distant reflection (which could potentially  lower the iron abundance), but the normalization of the distant reflector was orders of magnitude below the ionized reflection, and so did not improve the fit.


In order to get a better estimate of the corona electron temperature, we fit the spectrum with {\sc xillverCp} (Garcia et al., {\em in prep.}), which is a model of ionised reflection irradiated by a physical Comptonisation continuum, rather than a phenomenological cut-off powerlaw.  The thermal Comptonization model used is {\sc nthcomp} \citep{zdziarski96,zycki99}, which assumes a central spherical plasma surrounded by a geometrically thin disc. {\sc XillverCp} resulted in an equally good fit as those with {\sc xillver}, and the best-fit parameters were similar ($\chi^2/{\mathrm{dof}}=380/374=1.02$). See Fig.~\ref{ratio}d and Column~2 of Table~\ref{spec_params} for details. The electron temperature is constrained to be $15^{+2}_{-1}$~keV, which is similar to the temperature suggested by the cut-off powerlaw fit (assuming $E_{\mathrm{cut}}=2-3 kT_e$). Given the photon index ($\Gamma=2.32^{+0.02}_{-0.01}$) and electron temperature from {\sc nthcomp}, and using the equation for the photon index through solving the Kompaneets equation \citep{zdziarski96}, the corresponding optical depth is $\tau=2.7\pm0.2$.

Interestingly, the one parameter that does vary between {\sc xillverCp} and {\sc xillver} models is the iron abundance, which is close to solar values in model with a Comptonization continuum. This result might suggest that adding 2nd-order effects beyond the simplest reflection models could help solve the mystery of why some sources show statistically super-solar iron abundances.

\subsubsection{{\em Suzaku} + {\em NuSTAR} fits}

\begin{figure}
\includegraphics[width=\columnwidth]{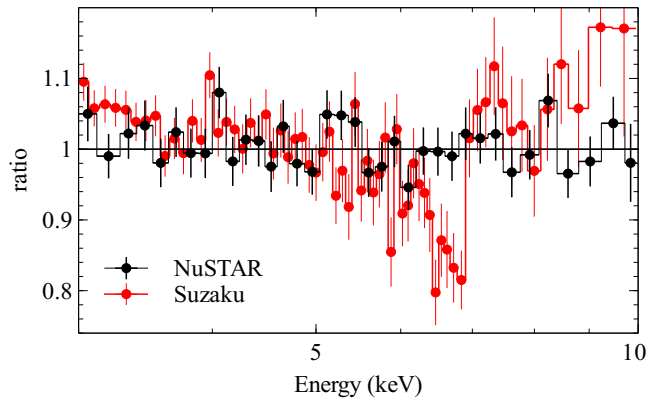}\\
\caption{The best fit {\sc xillver} model fit to the entire {\em NuSTAR} observation, now renormalized (i.e. letting photon index and normalization vary) to the 3-10~keV band of the 50~ks {\em Suzaku} observation and the 50~ks of the {\em NuSTAR} observation that was taken concurrently. While the unblurred reflection is a good fit to the {\em NuSTAR} data, the higher resolution {\em Suzaku} spectrum is not well fit. The ionized reflection model has a narrow core of the iron line that is not seen in the higher resolution {\em Suzaku} spectrum. This motivates the need for additional spectral complexities beyond a simple ionised reflection model, even though it is not required by the {\em NuSTAR} data alone (see Fig.~\ref{ratio}). }
\label{nusurat}
\end{figure}

Next we take the best fit ionized reflection model to the full {\em NuSTAR} spectrum, and apply it to the 50~ks {\em Suzaku} observation (concurrent with 50~ks of the {\em NuSTAR} spectrum).  In Fig.~\ref{nusurat} we show the residuals of these data to the best fit ionised reflection model to the full {\em NuSTAR} spectrum, allowing for differences in normalization and photon index in the two epochs. One can clearly see that while the simple ionised reflection is a good description of the {\em NuSTAR} spectrum, the {\em Suzaku} spectrum, which contains a similar number of counts in the iron~K band, has clear residuals ({\em Suzaku} has higher spectral resolution; 150~eV at 6~keV compared to 400~eV at 6~keV of {\em NuSTAR}).  The model contains a narrow peak at 6--7~keV that is not present in the data. We take this as evidence of further complexities in the iron line band.

Based on this evidence, we attempt to fit the entire spectrum from 1--50~keV with a {\sc relxilllp} model in order to further broaden the iron line.  This results in a fairly good fit ($\chi^2/{\mathrm{d.o.f}}=1299/1237=1.05$).  In order to fit the soft excess with the same relativistic reflection model as the iron line and Compton hump, the ionization parameter has decreased, and the iron abundance is pegged to its maximal value of 10 times solar value. Also, the black hole spin is required to be at its maximal value, and the coronal height is required to be very small.  This suggests that the extreme relativistic parameters are driven by the soft excess and not by the iron line or Compton hump.

\begin{figure*}
\includegraphics[width=\textwidth]{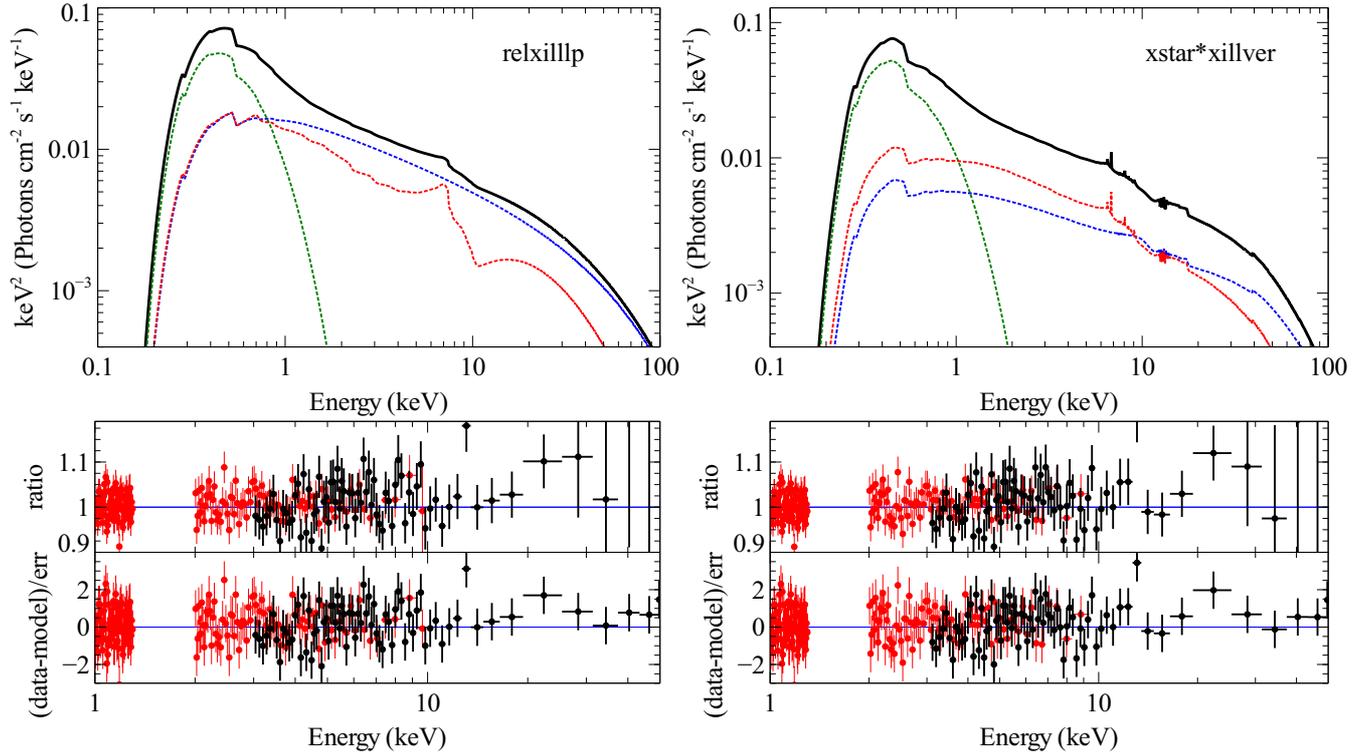}\\
\caption{{\em Left:}  The best-fit ({\sc relxilllp} and Bremsstrahlung model fit to the simultaneous {\em Suzaku} and {\em NuSTAR} (50~ks exposure) observations; $\chi^2/{\mathrm{d.o.f}}=1248/1235=1.01$). Below are the ratio of the data to the model and the $\Delta \chi^2$. {\em Right: } Same, but for the best-fit model of {\sc xillver} and Bremsstrahlung components through an ionised outflow ($\chi^2/{\mathrm{d.o.f}}=1246/1234=1.01$). See the last two columns of Table~\ref{spec_params} for parameters.}
\label{relxill_bremss_spec}
\end{figure*}

We find significant improvement by including an additional Bremsstrahlung component that contributes to the soft excess (see Column~3 of Table~\ref{spec_params}.  This leads to a $\Delta \chi^2=51$ for 2 additional degrees of freedom. The Bremsstrahlung component is largely phenomenological, though for a relatively low mass black at high accretion rate the disk is expected to be much denser than the standard assumption of $n_e=10^{15}$~cm$^{-3}$ (see Fig.~1 of \citealt{garcia16}). Recently, these authors showed that higher density discs (e.g. $n_e=10^{18}$~cm$^{-3}$) could lead to a significant increase in the free-free emission (by a factor of $\sim 5-10$).  Qualitatively similar results are found if we use an additional Comptonisation component (e.g. {\sc CompTT}; \citealt{titarchuk94}) in lieu of the Bremsstrahlung component (similar to \citealt{dewangan07}), which could be understood as the effect of bulk Comptonization due to turbulence in radiation pressure dominated accretion flows \citep{kaufman17}.  The blurred reflection and Bremsstrahlung model fit to the {\em Suzaku} and {\em NuSTAR} spectra can be seen in Fig.~\ref{relxill_bremss_spec}-{\em left}. One can see that only very modest relativsitic blurring is required, and near solar iron abundances provide very good fits.

Alternatively, the {\em Suzaku} and {\em NuSTAR} spectra can be equally well described by the ionised reflection and a Bremsstrahlung component through an ionised outflowing wind.  We use a photoionization table derived using the \emph{XSTAR} code version 2.2.1bn \citep{kallman01}. We consider a $\Gamma = 2$ power-law continuum and standard Solar abundances. We tested different broadenings due to turbulent velocities of 5,000~km~s$^{-1}$, 10,000~km~s$^{-1}$, and 30,000~km~s$^{-1}$. The latter is providing the best-fit and it is consistent with the large width observed for the absorption feature. This table is equivalent to the one used in \citet{tombesi15}. We show the results of this fit in Fig.~\ref{relxill_bremss_spec}, and in Column~4 of Table~\ref{spec_params}. This model also only requires solar iron abundance and a low cut-off energy.

The definitive results from our spectral fitting are that the reflection is very highly ionised for an AGN and that the coronal electron temperature is amongst the lowest yet observed.  Deeper observations are required to distinguish between relativistically blurred reflection and ionised reflection through a fast, outflowing wind. Of course, the solution could also be some combination of the two.

\section{Discussion}
\label{discuss}

Ark~564 is one of the most well-studied AGN in the X-ray band because it demonstrates the extremes of black hole accretion. Namely, it is one of the brightest AGN in soft X-rays, and is highly variable. It has a steeper spectrum than most Seyfert galaxies and a strong soft excess. In this paper, we have presented the first {\em NuSTAR} observations of this well-known object. In summary, our results are:
\begin{enumerate}
\item These {\em NuSTAR} observations show the best view of the iron~K complex in Ark~564, and reveal a higher equivalent width line than had been seen in previous, (potentially piled-up) {\em XMM-Newton} observations.
\item The feature above the iron line (due to a combination of the iron~K edge and photoelectric absorption smeared by Compton scattering in the hot upper layers of the disc) is seen up to $\sim 11$~keV. This is higher than other AGN observed with {\em NuSTAR}, and is similar to the profiles seen in black hole binaries, where the ionisation is typically larger.
\item The {\em NuSTAR} data are well described with a simple ionised reflection model with a high ionisation parameter, and do not require additional relativistic broadening or additional absorption from an ionised wind (though both of these models are adequate descriptions of the data).
\item Whether pure reflection or reflection plus absorption and regardless of a cut-off powerlaw continuum or a Comptonisation continuum, all of the attempted model fits described above require that the electron temperature of the corona is $\sim 15$~keV. This is one of the coolest temperature coronae measured to date.   
\item These {\em NuSTAR} data show the canonical hard lag as shown by the Fourier analysis and seen by eye in the data due to the fortuitous observation of a flare seen by both {\em Suzaku} and {\em NuSTAR}. This further suggests that we are seeing some direct emission from the continuum (rather than a highly obscured source). 
\end{enumerate}

In this section, we discuss the discovery a very low-temperature corona in Ark~564 in the context of previous studies of Ark~564 that suggest it is accreting at close to the Eddington limit and that it is an analog to a black hole binary in the luminous hard state.

\subsection{Low-temperature coronae in high-Eddington accretors}

In 2001 there was a large multiwavelength campaign of Ark~564 with {\em XMM-Newton}, {\em Chandra,} {\em ASCA}, {\em Hubble} and {\em FUSE}, which resulted in a broadband, simultaneous Spectral Energy Distribution \citep{romano04}. These authors find that the peak of the SED is in the Extreme UV/Soft X-rays and that Ark~564 has an intrinsic luminosity close to the Eddington limit.  This is consistent with the results from this analysis:  The average 2-10~keV luminosity over the entire observation (as derived from the best fit {\sc xillver} model fit to the {\em NuSTAR} and {\em Suzaku} data) is $3.9\times 10^{43}$~erg/s. Therefore, $L/L_{\mathrm{Edd}}=1.1$, assuming a bolometric correction of $9.2\pm4.5$ (from SED fitting; \citealt{vasudevan07}) and a black hole mass of $\log(M/M_{\odot})=6.4\pm0.5$ \citep{zhang06}. This high-Eddington luminosity is not completely surprising, as it has been suggested that NLS1s harbor low-mass black holes accreting at the Eddington limit (e.g. \citealt{boroson92}). 

\citet{pounds95} suggested that the steep spectrum in another NLS1 RE~J1034+39 could be explained in the context of super-Eddington accretion.  In this model the hard X-ray component had such a steep spectrum because the strong radiation field from the super-Eddington disc is cooling the corona more than in other sub-Eddington Seyferts.  In such a scenario, in addition to a steeper spectrum, we also expect a lower temperature corona (or a lower cut-off energy), which is what we see in Ark~564.  


Alternatively, \citet{laor14} describe a scenario (similar to \citealt{proga05}) in which UV line driven winds are launched in higher Eddington sources, causing an effective disc truncation radius that increases with Eddington ratio. They use this physical model to explain the observed constant 1000{\AA} turnover in AGN SEDs that is very weakly dependent on mass (in tension with standard thin-disc solutions).  \citet{laor14} suggest that the hot inner flow within the disc truncation radius could be Compton cooled, thus producing the observed hard X-ray emission.  Within this framework, higher Eddington sources would be cooled by relatively cooler seed photons, thus producing the lower temperature cut-off that is observed.  This result is consistent with our reflection fitting results for Ark~564, where only weak relativistic blurring is required, but could be in tension with observations of other NLS1s (e.g. 1H0707-495, IRAS~13224-3809), where extreme relativistic broadening is observed.

\subsection{Ark~564 as an analogue of the luminous-hard state in BHB}

Ark~564 is one of the brightest and most rapidly variable soft X-ray emitters, and therefore has been subject of many X-ray timing studies. The source was observed for several years with {\em RXTE}, {\em ASCA} and on short timescales with {\em XMM-Newton}, which allowed \citet{mchardy07} to perform power spectral density analysis over many decades in temporal frequency. The authors find that Ark~564 does not behave like any other Seyfert galaxy for which such long campaigns have been carried out. Most sources show a high frequency PSD slope of $\alpha \sim -2$, and a low-frequency slope of $\alpha \sim -1$ down to very low frequencies, which is reminiscent of the PSD of black hole binaries in the thermal-dominated state.  Ark~564, however, shows a second low-frequency break and a slope of $\alpha \sim 0$ at the lowest frequencies, very similar to BHB PSDs in all other states apart from the thermal-dominated state. Given the observed high-Eddington nature of this source, the authors suggest that Ark~564 is an AGN akin to the luminous hard state in BHBs.  

The hard X-ray spectral results presented in this paper are consistent with the interpretation that Ark~564 is in the luminous hard state.  We compare our spectral results with those of \citet{garcia15} using RXTE spectra of the BHB GX~339-4 in the hard state during an outburst.  These authors find that as the source goes into outburst the temperature of the corona decreases by an order of magnitude.  They also find that the ionization parameter in the reflection models is largest as the source hits its peak luminosity. These two correlations are consistent with our results of Ark~564 compared to other Seyferts observed with {\em NuSTAR}. The electron temperature of the corona is lower by a factor of two or more, and the ionization parameter is amongst the highest as well. This further supports the analogy between Ark~564 and black hole binaries in their luminous hard states.

While the spectral evidence found in this paper corroborates the temporal results of \citet{mchardy07} in suggesting that Ark~564 is an analogy to a black hole binary in its luminous hard state, we do emphasize that there are clear differences between AGN and BHB that may break the one-to-one link between the two mass regimes. For example, it has long been understood that AGN discs are more radiation pressure dominated than those in BHBs \citep{shakura73}, and, as discussed above, UV line driven winds can be important in cooler AGN discs, which can affect the AGN disc structure.

\subsection{Implications for coronal parameters}

Recently, \citet{fabian15} compiled a sample of all the high-energy cut-offs observed with {\em NuSTAR} and populated these sources on the compactness-temperature ($\ell-\Theta$) plane, where $\Theta=kT_{e}/m_{e} c^2$ is the coronal electron temperature normalized by the electron rest energy and $\ell=(L/R)(\sigma_{T}/m_{e} c^3)$ is the dimensionless compactness parameter \citep{guilbert83}. \citet{fabian15} defined $L$ as the luminosity of the powerlaw component from 0.1--200~keV and $R$ as the radius of the corona estimated by X-ray reverberation estimates, where available, or simply assumed to be $10 r_{\mathrm{g}}$, where reverberation measurements were not possible. 

Theoretical constraints limit the allowed parameter space on the $\ell-\Theta$ plane, the most important of which is the pair-production limit.  In compact and high-temperature coronae, electrons lose a significant fraction of their energy to photons, which can become so energetic that photon-photon collisions produce electron-positron pairs (i.e. when the product of the photon energies exceeds 2 (in units of $m_e c^2$).  Thus any additional energy input will lead to an increased number of pairs, but does not increase the source temperature.  Nearly all sources observed up to that point by {\em NuSTAR} lay just below the pair-production limit for thermal Comptonisation, suggesting that these coronae are pair-dominated plasmas.  

We compare our results of Ark~564 with the results of \citet{fabian15}.  Our modelling of the spectrum with {\sc nthcomp} irradiating an ionized accretion disc, found an electron temperature of 15~keV (or $\Theta=0.03$).  Following \citet{fabian15}, we measure the luminosity of the powerlaw component from 0.1--200~keV to be $1.1\times 10^{44}$~erg/s.  Ark~564 does have a reverberation lag measurement, and so we use the height of the corona constrained by modelling the reverberation lags to estimate the radius of a spherical corona ($R=4.6~r_{\mathrm{g}}$; \citealt{emm14}). This leads to a compactness of $\ell=1140$. Ark~564 lays well below the thermal pair-production limit (see Fig.~2 of \citealt{fabian15}), and thus, assuming a thermal Comptonisation model, we find that the corona in this source is not a pair-dominated plasma, but rather is mostly normal matter.   

However, recently, \citet{fabian17} re-examined the case of hybrid coronae \citep{zdziarski93}, where the plasma contains both thermal and non-thermal particles, as might be expected for a highly magnetized corona powered by the dissipation of magnetic energy. In such a compact, highly magnetized corona, heating and cooling are so intense that the electrons might not have time to thermalize before inverse Compton cooling. If a small fraction of electrons follow a non-thermal distribution that exceeds MeV temperatures, this can cause run-away pair production. Cooled pairs could share their available energy, thus reducing the mean energy per particle and reducing the temperature of the thermal population. So unless we have the sensitivity to probe the hard non-thermal tail of the electron distribution, a low cut-off energy could either be interpreted as a thermal plasma that is not pair dominated or as a pair-dominated hybrid plasma. Further deep hard X-ray observations are required to distinguish these two scenarios.

Ark~564 lays definitively below the thermal pair production limit and very close to the limit where electron-electron coupling timescale is equal to the timescale for Compton cooling. Indeed the only sources from \citet{fabian15} in that part of the parameter space are two black hole binaries, GRS 1915+105 \citep{miller13} and GRS 1739-278 \citet{miller15}, observed during the luminous phases of their low/hard states. This is consistent with the idea that Ark~564 is an AGN analogue to the luminous hard state in black hole binaries. 

\section{Conclusions}

We have presented the analysis of the new {\em NuSTAR} observation of the steep spectrum Narrow-Line Seyfert 1 galaxy Ark~564.  This source is an order of magnitude brighter than other steep spectrum NLS1s measured by {\em NuSTAR}, which has allowed for the first determination of the cut-off energy in these types of sources. We find a very low temperature of $15\pm2$~keV, amongst the lowest measured by {\em NuSTAR} of any Seyfert galaxy.  Steep spectrum NLS1s are thought to be accreting at close to the Eddington limit, and so further observations of Comptonisation at these most extreme limits are important for understanding the connection between the accretion disc and its corona.

\section*{Acknowledgements}

EK thanks Julian Malzac, Pierre-Olivier Petrucci, Andrea Marinucci and the participants of the FERO8 meeting in Moravia, Czech Republic for interesting discussions on accretion disc coronae. EK thanks the NuSTAR GO Program for support under grant NNX15AV26G. EK thanks the Hubble Fellowship Program. Support for Program number HST-HF2-51360.001-A was provided by NASA through a Hubble Fellowship grant from the Space Telescope Science Institute, which is operated by the Association of Universities for Research in Astronomy, Incorporated, under NASA contract NAS5-26555. ACF acknowledges ERC Advanced Grant 340442. CSR thanks support from NASA under grant NNX17AF29G.

\bibliography{references}   
\bibliographystyle{mnras.bst}      

\end{document}